\documentclass[aps, prd, showpacs, twocolumn, superscriptaddress, nofootinbib]{revtex4-1}
\usepackage{graphicx}
\usepackage{amsmath}
\usepackage{gensymb}
\usepackage{xcolor}

\renewcommand{\v}[1]{\ensuremath{{\mathbf{#1}}}}
\newcommand{\vpq}{\ensuremath{\vert \v{p} + \v{q} \vert}}
\newcommand{\neff}{\ensuremath{N_{\rm eff}}}
\begin{document}

\begin{abstract}
Electron-positron annihilation largely occurs in local thermal and chemical equilibrium after the neutrinos fall out of thermal equilibrium and during the Big Bang Nucleosynthesis (BBN) epoch. The effects of this process are evident in BBN yields as well as the relativistic degrees of freedom. We self-consistently calculate the collision integral for electron-positron creation and annihilation using the Klein-Nishina amplitude and appropriate statistical factors for Fermi-blocking and Bose-enhancement.  Our calculations suggest that this annihilation freezes out when the photon-electron-positron-baryon plasma temperature is approximately 16 keV, after which its rate drops below the Hubble rate. In the temperature regime near 16 keV, we break the assumption of chemical equilibrium between the electrons, positrons, and photons to independently calculate the evolution of the chemical potentials of the electrons and positrons while computing the associated collision integrals at every time step. We find that the electron and positron chemical potentials deviate from the case with chemical equilibrium. While our results do not affect the interpretation of precision cosmological measurements in elucidating the standard cosmological model, these out of equilibrium effects may be important for testing physics beyond the standard model.

\end{abstract}

\title{Electron-Positron Annihilation Freeze-Out in the Early Universe}
\author{Luke C.\ Thomas}
\affiliation{Department of Physics and Biophysics, University of San Diego, San Diego, CA 92110}
\author{Ted Dezen}
\affiliation{Department of Physics and Biophysics, University of San Diego, San Diego, CA 92110}
\author{Evan B.\ Grohs}
\affiliation{Department of Physics, University of California, Berkeley, Berkeley, CA 94720}
\author{Chad T.\ Kishimoto}
\affiliation{Department of Physics and Biophysics, University of San Diego, San Diego, CA 92110}
\date{eventually}
\affiliation{Center for Astrophysics and Space Sciences, University of California, San Diego, La Jolla, CA 92093}
\date{\today}
\maketitle

\section{Introduction}
The chronology of the early universe is the story of a hot, dense, and entropy-dominated plasma that may have, at one time, included thermal populations of all particles in the Standard Model (the so-called ``primordial soup'').  While particle populations are well described by thermal and chemical equilibria, the standard cosmological model incorporates phase transitions that change the nature of the fundamental forces in the Standard Model as well as epochs where various processes freeze-out of equilibrium.  Freeze-out from equilibrium is not an instantaneous process, resulting in deviations of particle phase-space distributions as the approximations of local thermal and chemical equilibrium become invalid \cite{burst,dp16,man05,dhpprs02}.

Upcoming high-precision cosmological observations \cite{CMBS4,EELT-HIRES} introduce the promise of the early universe as a testing ground for the Standard Model and the standard cosmological model. Out-of-equilibrium effects that occur when freeze-out is not instantaneous are an especially promising regime for revealing Beyond Standard Model (BSM) physics \cite{gfkp15,cddks18,fppr14,bjw16,mcd18,jp09}.  

The well-studied epochs of weak decoupling and weak freeze-out occur when the temperature of the photon-electron-positron-baryon plasma, $T$, is approximately an MeV \cite{kt90}.  Weak decoupling ($T \sim 3~{\rm MeV}$) occurs when neutrinos no longer efficiently exchange energy with the plasma so they are no longer in thermal equilibrium.  Weak freeze-out ($T \sim 0.7~{\rm MeV}$) occurs when the charged current lepton-nucleon processes inefficiently inter-convert protons to neutrons, freezing-out the neutron-to-proton ratio as they fall out of chemical equilibrium with the leptons.  These processes occur over many Hubble times (expansion timescales), requiring a self-consistent treatment of the Boltzmann evolution of neutrino distribution functions and a nuclear network to describe with the consequent Big Bang Nucleosynthesis (BBN) \cite{burst}.  The results of these computationally-intensive calculations may be observable in the energy density of neutrinos (as parameterized by $N_{\rm eff}$) and in primordial abundances as the next generation of cosmic microwave background (CMB) observatories and thirty-meter telescopes come on-line \cite{CMBS4,EELT-HIRES}.

Careful analysis of this epoch reveals a variety of Standard Model processes that affect the value of \neff{} and BBN yields {\it because} weak decoupling and weak freeze-out are not instantaneous.  The better known of these effects is the increase of \neff{} from its standard, instantaneous decoupling value of 3 to the accepted value of 3.046 due to a variety of effects including electron-positron annihilation, out-of-equilibrium neutrino up-scattering off the photon-electron-positron-baryon plasma, and finite temperature QED radiative corrections to the in-medium electron and photon masses \cite{burst,byr15,dhs97,man02}.  Other work has looked into the effects of neutrino quantum kinetic behavior \cite{man05,dp16} and non-standard cosmological models including the introduction of non-zero lepton numbers \cite{gfkp17}.  BBN yields are also affected through their influence on neutrino distribution functions and the time-temperature relationship throughout the BBN epoch.

These in-depth studies are necessary to disentangle out-of-equilibrium Standard Model effects from BSM physics, both of which may have an observable signal in \neff{} and the BBN yields.  During the BBN epoch, electron-positron creation becomes kinematically suppressed compared to annihilation, driving the equilibrium process -- $e^- e^+ \rightleftharpoons \gamma \gamma$ -- to the right, resulting in the number densities of electrons and positrons plummeting.  In the standard cosmological model, BBN yields and \neff{} are calculated as if this process always occurs in local thermal and chemical equilibrium \cite{kawano}.

The goal of this work is to closely examine the process of electron-positron annihilation  as it freezes out from its local chemical equilibrium. We will determine if and when the electron and positron number densities freeze-out, investigate the consequent effects on cosmological observables, and speculate on whether the effects of freeze-out should affect the standard BBN calculations.  To do so, we calculate the collision integral for electron-positron annihilation using the Klein-Nishina amplitude with the appropriate Fermi-blocking/Bose-enhancing statistical factors.  Appendix \ref{sec:appb} details our methodology to self-consistently and efficiently calculate these integrals using methods similar to those in Ref.\ \cite{burst}.

Throughout this work we use the convention that $\hbar = c = k_B = 1$.  In Section \ref{sec:chemeq}, we use physical principles, including chemical equilibrium between the electrons and positrons, to introduce the standard cosmological computation to evolve the electron and positron distributions.  Next, in Section \ref{sec:coll}, we calculate the collision integral associated with electron-positron annihilation to infer when freeze-out occurs. We subsequently use the collision integral to separately evolve the electron and positron distributions without the assumption that the charged leptons maintain chemical equilibrium in Section \ref{sec:nochemeq}.  In Section \ref{sec:conclusion} we will discuss the results and present conclusions.  Appendix \ref{sec:appa} details the equations of motion derived and solved in this work.

\section{Chemical equilibrium} \label{sec:chemeq}

As evidenced by the cosmic microwave background, the primordial plasma is well described by local thermodynamic equilibrium. Scattering between photons, electrons, positrons, and baryons remains fast compared to the Hubble rate, maintaining thermal distributions well into the BBN epoch. This means the photons, electrons, positrons, and baryons share a single plasma temperature $T$, and the electron/positron distributions are Fermi-Dirac. When further assuming the process $e^- e^+ \rightleftharpoons \gamma \gamma$ to be in local chemical equilibrium, the chemical potentials of the involved particles maintain $\mu_{e^-} + \mu_{e^-} - 2\mu_\gamma = 0$, stipulating $\mu_{e^+} = -\mu_{e^-}$. For chemical potentials $\mu_{e^\pm}$, we define the electron and positron degeneracy parameters $\eta^- \equiv \mu_{e^-}/T$ and $\eta^+ \equiv \mu_{e^+}/T$. Chemical equilibrium is implemented by  setting $\eta^-=\eta$ and $\eta^+=-\eta$. 

Given local thermodynamic and chemical equilibrium, the electron and positron momentum/energy distributions are
\begin{align}
	f_{e^-}(p) & =\frac{1}{e^{E_p/T-\eta}+1} \\
	f_{e^+}(p)& =\frac{1}{e^{E_p/T+\eta}+1},
\end{align}
where $E_p=\sqrt{p^2+m_e^2}$. The evolution of the electron and positron distributions is encapsulated in the evolution of $T$ and $\eta$. To evolve these quantities, we introduce two physical principles -- conservation of comoving entropy, and the effects of weak interactions on the relative number of electrons and positrons.

The early universe is homogeneous and isotropic, precluding spatial heat flows. In the absence of out-of-equilibrium decays, the total entropy in a comoving volume is conserved,
\begin{equation}
    \frac{d}{dt}[sa^3]=0.
    \label{eq:entropy_conservation}
\end{equation}
Here $a$ is the scale factor in the Friedmann-Lema\^{i}tre-Robertson-Walker metric and entropy density is
\begin{equation}
    s=\frac{\rho+P}{T} - \eta n_{e^-} + \eta n_{e^+}.
\end{equation}
The quantities $\rho$ and $P$ are total energy density and pressure of all particles in the plasma, while electron and positron number densities are denoted $n_{e^\pm}$.  The scale factor evolves according to the Friedmann equation,
\begin{equation}
    \frac{1}{a}\frac{da}{dt} = \sqrt{\frac{8\pi}{3m_{pl}^2}\rho},
    \label{eq:friedmann1}
\end{equation}
where $m_{pl}$ is the Planck mass.

The number densities of electrons and positrons are affected by the expansion of the universe, creation and annihilation, and charged-current weak interactions. The product $na^3$ is proportional to the number of particles in a comoving volume, which remains unchanged by the expansion of the universe. Creation and annihilation rates are fast compared to the weak rates, but are computationally expensive to calculate. In order to avoid the need to self-consistently calculate these rates with the evolution, we introduce the quantity $(n_{e^-}-n_{e^+})a^3$, which is insensitive to both the expansion of the universe and electromagnetic processes.  Electromagnetic processes preserve the relative number of electrons and positrons, so the evolution of this quantity does not depend on the collision integral, even when the assumption of chemical equilibrium is broken.

Weak interactions that inter-convert neutrons and protons are the only processes that cause $(n_{e^-}-n_{e^+})a^3$ to change. During the BBN epoch, these processes are
\begin{align}
    \label{eq:weak_interaction1}
    \nu_e+n&\rightleftharpoons p+e^- \\
    \label{eq:weak_interaction2}
    e^++n&\rightleftharpoons p+\bar{\nu}_e\\
    \label{eq:weak_interaction3}
    n&\rightleftharpoons p+\bar{\nu}_e+e^-.
\end{align}
When the above reactions proceed to the right, $(n_{e^-}-n_{e^+})a^3$ increases, while the reverse reactions cause it to decrease. The rate of change of this quantity is
\begin{widetext}
\begin{equation}
    \frac{d}{dt}\big[(n_{e^-}-n_{e^+})a^3\big]=n_na^3(\lambda_{\nu_en}+\lambda_{e^+n}+\lambda_n)-n_pa^3(\lambda_{e^-p}+\lambda_{\bar{\nu}_ep}+\lambda_{\bar{\nu}_ee^-p}).
    \label{dndt_weak1}
\end{equation}
The rates $\lambda_i$, with $i$ indicating the reactants, are computed in Ref.\ \cite{gf16}. Rather than directly computing the number densities of neutrons $(n_n)$ and protons $(n_p)$, we can recast Eq. (\ref{dndt_weak1}) in terms of the electron fraction
\begin{equation}
    Y_e=\frac{n_p}{n_b}=\frac{n_{e^-}-n_{e^+}}{n_b},
    \label{eq:yedef}
\end{equation}
resulting in
\begin{equation}
	\frac{1}{n_ba^3}\frac{d}{dt}\left[ (n_{e^-}-n_{e^+}) a^3 \right] =(1-Y_e)(\lambda_{\nu_en}+\lambda_{e^+n}+\lambda_n)-Y_e(\lambda_{e^-p}+\lambda_{\bar{\nu}_ep}+\lambda_{\bar{\nu}_ee^-p}),
	\label{eq:dndt_weak2}
\end{equation}
where $n_b$ is the number density of baryons, and the quantity $n_b a^3$ is constant.  The right-hand-side comes from the rate of change of $Y_e$ \cite{gf16},
\begin{equation}
    \frac{d Y_e}{dt} = (1-Y_e)(\lambda_{\nu_en}+\lambda_{e^+n}+\lambda_n)-Y_e(\lambda_{e^-p}+\lambda_{\bar{\nu}_ep}+\lambda_{\bar{\nu}_ee^-p}) ,
    \label{eq:dyedteqm}
\end{equation}
which accounts for the production and destruction of electrons and positrons by the weak interactions.  
\end{widetext}

Together, the conservation of comoving entropy (Eq. \ref{eq:entropy_conservation}), the evolution of the difference in comoving numbers of electrons and positrons due to weak interactions (Eq. \ref{eq:dndt_weak2}), the evolution of $Y_e$ (Eq.\ \ref{eq:dyedteqm}), and the Friedmann equation (Eq.\ \ref{eq:friedmann1}) form a system of coupled differential equations for dependent variables $T$, $\eta$, $Y_e$, and $a$ (see Section 1 of Appendix \ref{sec:appa}).  It should be noted that the definition of $Y_e$, Eq.\ (\ref{eq:yedef}), is an integral relationship between all four dependent variables, so while the four aforementioned differential equations are not independent, we find it more efficient to simultaneously use them to evolve the four variables.

We elect an initial plasma temperature $T_i=10\,{\rm MeV}$, sufficiently high that we are assured that the electron-positron annihilation rate is much greater than the Hubble rate. At this temperature, the weak interactions (Eqs. \ref{eq:weak_interaction1}-\ref{eq:weak_interaction3}) are in chemical equilibrium so that the initial electron fraction is \cite{kt90}
\begin{equation}
    Y_{e,i}=\bigg[\exp\bigg(-\frac{m_n-m_p}{T_i}+\eta_i\bigg)+1\bigg]^{-1}.
    \label{eq:initial_Ye}
\end{equation}
The initial degeneracy parameter, $\eta_i$, is chosen to be consistent with the observed baryon-to-photon ratio from CMB observations \cite{planck},
\begin{equation}
    \frac{n_{b,f}}{n_{\gamma,f}} = \frac{n_{b,f}}{\frac{2\zeta(3)}{\pi^2}T_f^3} = 6.1\times10^{-10},
\end{equation}
where $\zeta(3)\approx1.20206$ and the subscript $f$ indicates the values at the end of the calculation. (Here, we assume there is no BSM physics that changes the baryon-to-photon rates between the keV scale and recombination \cite{gfkp15}.) Since $n_b a^3$ is conserved, the initial baryon number density is
\begin{equation}
    n_{b_i}=(6.1\times10^{-10})\frac{2\zeta(3)}{\pi^2}\bigg(\frac{a_fT_f}{a_iT_i}\bigg)^3T_i^3.
\end{equation}
In the standard cosmological model \cite{kt90,gf17},
\begin{equation}
    T_f a_f = \bigg(\frac{11}{4}\bigg)^{\frac{1}{3}} T_i a_i,
\end{equation}
which accounts for the heating of the plasma when the rest mass energy in electrons and positrons is liberated by annihilation. The initial baryon number can be related to the initial degeneracy parameter through integration of the Fermi-Dirac distribution, 
\begin{equation}
    n_{b,i}=\frac{1}{Y_{e,i}}(n_{e^-}-n_{e^+})_i\approx\frac{1}{Y_{e,i}}\frac{1}{3}\eta_iT_i^3,
    \label{eq:initial_eta}
\end{equation}
assuming that $T\gg m_e$ and $\eta_i \ll 1$.  Eqs.\ (\ref{eq:initial_Ye}-\ref{eq:initial_eta}) produce the initial conditions to evolve $T$, $\eta$, $Y_e$, and $a$ through the epoch where annihilation is important.

\begin{figure}[]
    \centering
    \includegraphics[width=0.9\columnwidth]{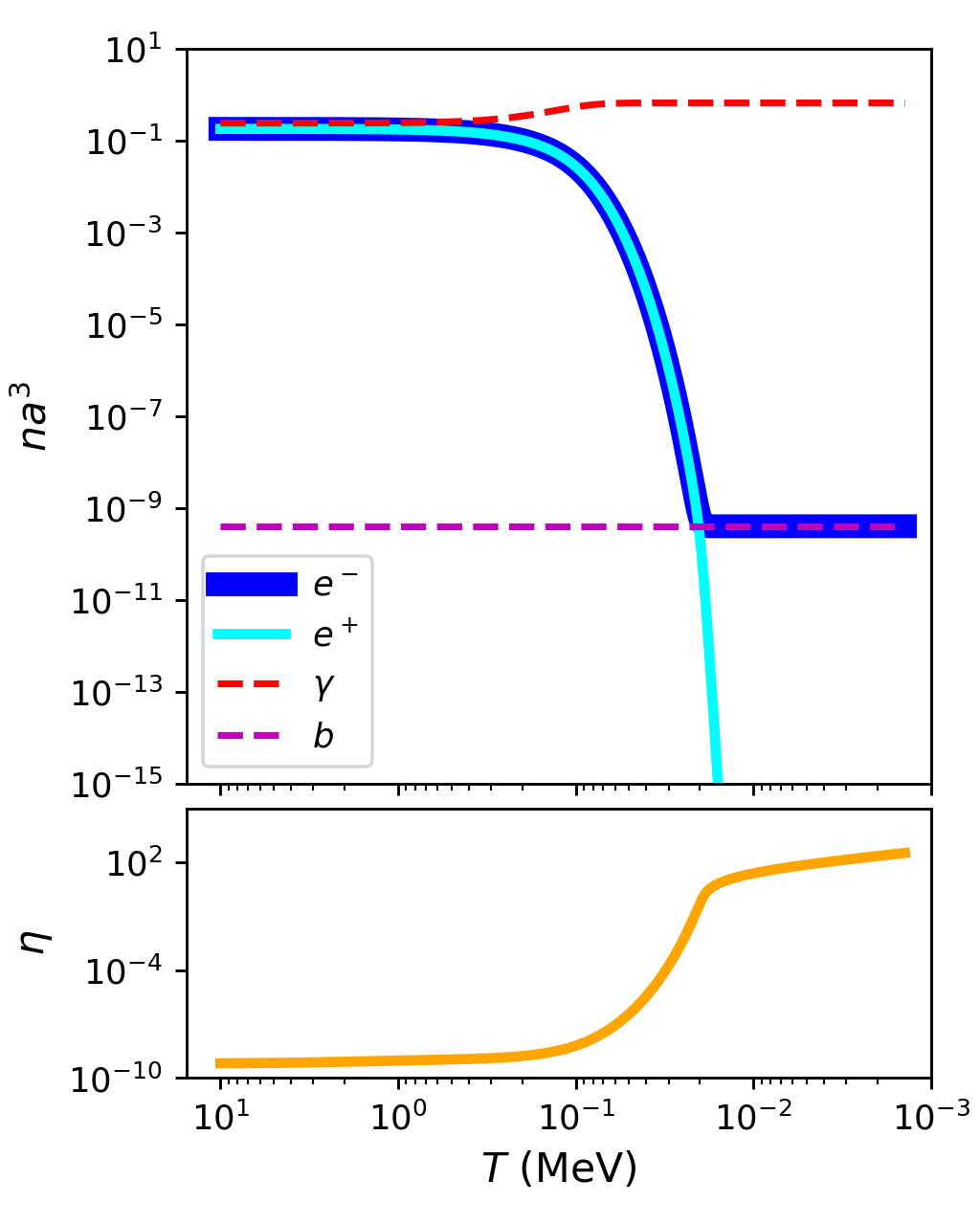}
    \caption{Top panel:  number of particles in a comoving volume for electrons (thick solid curve), positrons (thin solid curve), photons (upper dashed curve), and baryons (lower dashed curve) as a function of plasma temperature.  Bottom panel:  electron degeneracy parameter, $\eta$ (with $\eta^- = \eta$ and $\eta^+ = - \eta$).}
    \label{fig:number_densities}
\end{figure}
Figure \ref{fig:number_densities} presents the results of this chemical equilibrium solution. The number of particles in a comoving volume for each species in the photon-electron-positron-baryon plasma is plotted as a function of decreasing plasma temperature. When the plasma temperature drops below about an MeV, pair production becomes energetically suppressed, causing the numbers of electrons and positrons to fall precipitously, with positrons tracking electrons until electron numbers abruptly level off and positron numbers continue to plummet.  As seen in Fig.\ \ref{fig:number_densities}, this difference is driven by rapid growth in $\eta$.

Local chemical equilibrium should hold so long as the creation/annihilation rates for $e^-e^+\rightleftharpoons\gamma\gamma$ remain fast relative to the Hubble rate. The rapid decrease in electron and positron numbers causes the equilibrium creation/annihilation rate to similarly fall off.  The following section utilizes the equilibrium solution to calculate the annihilation rate to assess the possibility of annihilation freeze-out.

One minor issue with this calculation is the lack of a BBN nuclear network calculation.  BBN would ``protect'' neutrons from undergoing free neutron decay by locking up the free neutrons primarily in ${}^4{\rm He}$ nuclei.  In the standard cosmological model, this occurs at $T \sim 100~{\rm keV}$ and results in $Y_e \approx 0.88$ at late times (while in our calculation $Y_e \rightarrow 1$ as free neutron decay eventually converts every neutron into a proton).  To ensure that this approximation has negligible effects on our results, we performed the same calculation with a constant $Y_e$ equal to its initial value throughout the evolution, and found no significant changes in the conclusions.  Having bracketed the solution that includes BBN, we expect that self-consistently following the nuclei through BBN will not affect the conclusions of this work.

\section{Annihilation rates} \label{sec:coll}

To determine when the annihilation process $e^-e^+\rightarrow\gamma\gamma$ freezes out, we compute the fractional rate of change of the number density of electrons and positrons, due solely to annihilation, 
\begin{equation}
\mathcal{R}_{e^\pm}\equiv\frac{1}{n_{e^{\pm}}} \bigg\vert \frac{dn_{e^\pm}}{dt} \bigg\vert_{e^-e^+\rightarrow\gamma\gamma}.
\end{equation}
The time derivative of the number densities can be calculated directly from the collision integral,
\begin{align}
    \frac{dn_{e^\pm}}{dt}\bigg\vert_{e^-e^+\rightarrow\gamma\gamma} & =  \frac{d}{dt} \int_0^\infty \frac{p^2dp}{2\pi^2} \,f_{e^\pm}(p) \nonumber \\
    & = - \int_0^\infty \frac{p^2dp}{2\pi^2} \,C_a(p).
    \label{eq:dndt_annihilation}
\end{align}
Here, $C_a$ is the collision integral for the annihilation pathway,
\begin{equation}
    \frac{df_{e^\pm}}{dt}\bigg\vert_{e^-e^+\rightarrow\gamma\gamma}= -C_a(p),
\end{equation}
which is one of many terms in the Boltzmann equation for the rate of change of $f_{e^\pm}$.  Appendix \ref{sec:appb} outlines the computation of the collision integral.

\begin{figure}[]
    \centering
    \includegraphics[width=0.95\columnwidth]{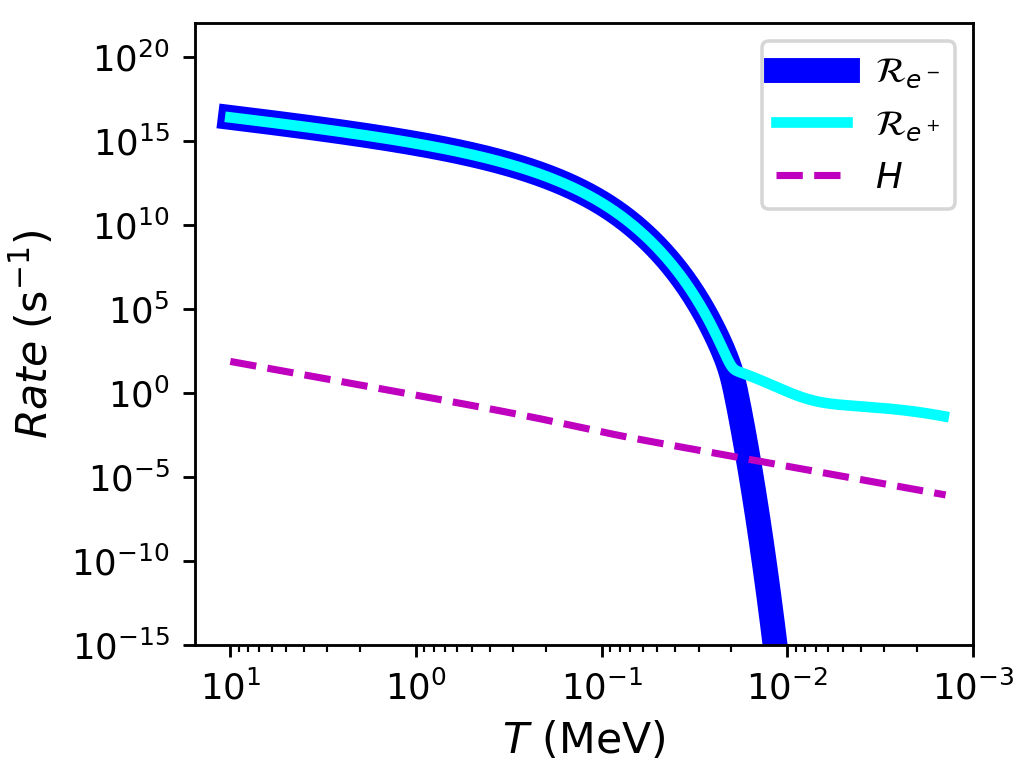}
    \caption{Annihilation rates for electrons ($\mathcal{R}_{e^-}$, thick solid line) and positrons ($\mathcal{R}_{e^+}$, thin solid line) as a function of decreasing plasma temperature.  Also plotted is the Hubble rate (dashed line) for comparison.  Each of these rates are calculated using the results from the chemical equilibrium solution.}
    \label{fig:annihilation_rate}
\end{figure}

Figure \ref{fig:annihilation_rate} presents $\mathcal{R}_{e^\pm}$ calculated using $T$, $\eta$, $Y_e$ and $a$ from the chemical equilibrium solution.  Due to the paucity of positrons, $\mathcal{R}_{e^-}$ drops below the Hubble rate at 16 keV, so at this time the universe expands much faster than the rate at which electrons annihilate with positrons.  However, $\mathcal{R}_{e^+}$ remains many orders of magnitude faster than the Hubble rate as it is much more likely for a positron to annihilate on the significantly more abundant electrons, so essentially every positron will annihilate with an electron.  

\section{Out of equilibrium} \label{sec:nochemeq}

With annihilation freeze-out estimated to occur at 16 keV, we lift the assumption of local chemical equilibrium and reexamine this period more closely.  Electrons and positrons continue to efficiently scatter with the plasma, so their distributions are still Fermi-Dirac,
\begin{equation}
    f_{e^\pm}(p)=\frac{1}{e^{E_p/T-\eta^\pm}+1},
\end{equation}
but allowing for deviations from chemical equilibrium implies that the electron and positron degeneracy parameters evolve separately.  With independent electron and positron degeneracy parameters, we need to re-evaluate the four coupled equations of motion used in the chemical equilibrium solution, Eqs.\ (\ref{eq:entropy_conservation}, \ref{eq:friedmann1}, \ref{eq:dndt_weak2}, \ref{eq:dyedteqm}), to account for the separate degeneracy parameters (see Section 2 in Appendix \ref{sec:appa}).  In addition, we introduce the time evolution of the sum of electron and positron numbers in a comoving volume, $(n_{e^-}+n_{e^+})a^3$, which is modified by weak interactions as well as creation and annihilation processes:
\begin{widetext}
\begin{equation}
    \frac{d}{dt}\big[(n_{e^-}+n_{e^+})a^3\big] = n_na^3(\lambda_{\nu_en}-\lambda_{e^+n}+\lambda_n) + n_pa^3(-\lambda_{e^-p}+\lambda_{\bar{\nu}_ep}-\lambda_{\bar{\nu}_ee^-p}) + 2a^3\mathcal{N}.
    \label{eq:dndt_sum}
\end{equation}
\end{widetext}
The first two terms on the right-hand-side account for the creation and absorption of electrons and positrons in the weak interactions, and $\mathcal{N}$ accounts for out-of-equilibrium effects from creation/annihilation,
\begin{equation}
    \mathcal{N} = \frac{dn_{e^-}}{dt}\bigg\vert_{e^-e^+\rightleftharpoons\gamma\gamma} = \frac{dn_{e^+}}{dt}\bigg\vert_{e^-e^+\rightleftharpoons\gamma\gamma} .
\end{equation}

To determine $\mathcal{N}$, we self-consistently calculate the collision integral for creation/annihilation, $C(p)$,
\begin{equation}
    \frac{df_{e^\pm}}{dt}\bigg\vert_{e^-e^+\rightleftharpoons\gamma\gamma} = C_c (p) - C_a (p) \equiv C(p) ,
\end{equation}
where $C_a (p)$ is the collision integral for the annihilation pathway as in the previous section, and $C_c (p)$ is for the creation pathway.  It follows that 
\begin{equation}
    \mathcal{N} = \int_0^\infty \frac{p^2dp}{2\pi^2} C(p) .
    \label{eq:dndt_netrate}
\end{equation}
Further details on the collision integral calculations can be found in Appendix \ref{sec:appb}.

Together, Eqs.\ (\ref{eq:entropy_conservation}, \ref{eq:friedmann1}, \ref{eq:dndt_weak2}, \ref{eq:dyedteqm},  \ref{eq:dndt_sum}) yield a system of coupled differential equations for dependent variables $T$, $\eta^-$, $\eta^+$, $Y_e$, and $a$.  As with the chemical equilibrium solution, the definition of $Y_e$, Eq.\ (\ref{eq:yedef}), is an integral relationship between all the dependent variables. While this statement again means that the differential equations are not independent, we simultaneously solve the five differential equations for the five variables, as we did in the chemical equilibrium solution.

Figure \ref{fig:annihilation_rate} shows that the creation and annihilation rates are many orders of magnitude greater than the Hubble rate for $T > 16~{\rm keV}$, which would require far too many time steps to integrate these equations from $T = 10~{\rm MeV}$ as we did with the chemical equilibrium solution.  However, because we expect chemical equilibrium to hold for $T > 16~{\rm keV}$, we can use the chemical equilibrium solution to determine initial conditions for this non-equilibrium solution.

\begin{figure}[]
    \centering
    \includegraphics[width=0.9\columnwidth]{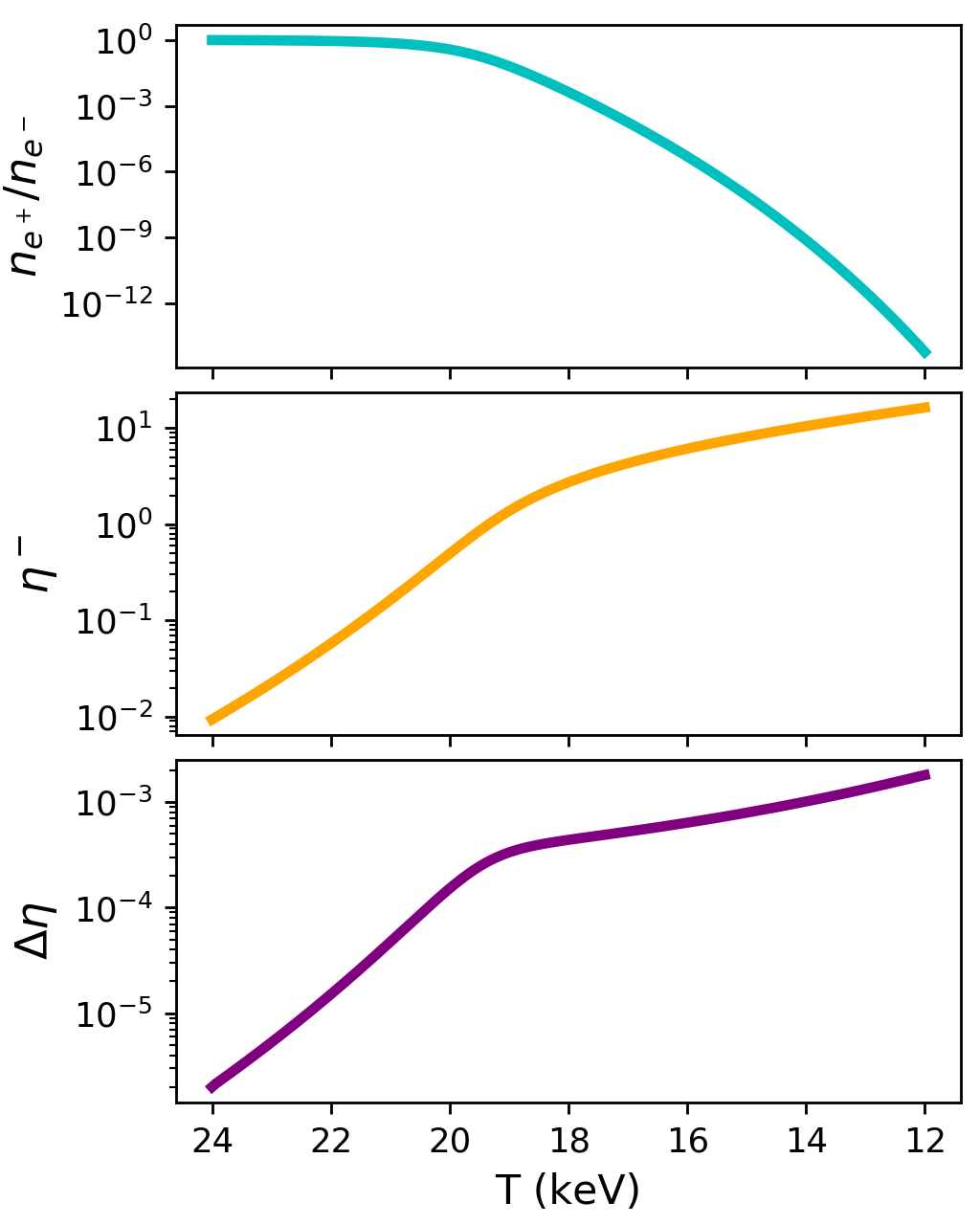}
    \caption{Results from the non-equilibrium solution using the chemical equilibrium solution as the initial condition at $T = 24~{\rm keV}$.  From top to bottom, the plots show the positron-to-electron ratio ($n_{e^+}/n_{e^-}$), electron degeneracy parameter ($\eta^-$), and the deviation of the electron and positron degeneracy parameters from equilibrium, $\Delta \eta=\eta^-+\eta^+$, for plasma temperatures from 24 keV to 12 keV.}
    \label{fig:Delta_eta}
\end{figure}

Figure \ref{fig:Delta_eta} presents the solution to these fully non-equilibrium conditions between 24 and 12 keV, lasting about a Hubble time and bracketing the period of interest.  The initial conditions for $Y_e$ and $a$ are taken from the chemical equilibrium solution at $T = 24~{\rm keV}$, and the initial degeneracy parameters for electrons and positrons are derived from $\eta$ in the chemical equilibrium solution at $T = 24~{\rm keV}$ with $\eta^- = \eta$ and $\eta^+ = - \eta$.  

Deviation from $\eta^+=-\eta^-$ indicates a shift away from local chemical equilibrium, so the quantity $\Delta\eta=\eta^-+\eta^+$ parameterizes the magnitude of any such out-of-equilibrium effect.  The initial conditions set the initial value of $\Delta \eta = 0$, but the non-equilibrium solution immediately deviates from this initial condition.  Throughout the evolution, $\Delta \eta \ll \eta^-$, and the electron degeneracy parameter in the non-equilibrium solution ($\eta^-$) differs from the equilibrium solution ($\eta$ in Figure \ref{fig:number_densities}) by an amount on the order of $\Delta \eta$.  As a result, the positron-to-electron ratio and the electron chemical potential as shown in Figure \ref{fig:Delta_eta} slightly differ from the corresponding values in the chemical equilibrium solution with a fractional difference on the order of $\Delta \eta$.

To test the validity of taking the initial conditions directly from the chemical equilibrium solution, we repeated the process of determining initial conditions at 25 keV, then evolved the non-equilibrium equations.  Once again, the initial conditions are chosen such that the initial $\Delta \eta = 0$ and the non-equilibrium solution immediately deviates from this value.  We found that this solution was consistent with the results presented in Fig.\ \ref{fig:Delta_eta} for $24~{\rm keV} < T < 12~{\rm keV}$.

There are two possible observable consequences of this out-of-equilibrium process:  BBN yields and \neff.  The light element abundances formed in BBN are sensitive to the time-temperature relationship and the neutron-to-proton ratio during the BBN epoch ($\sim 100~{\rm keV}$), both of which are influenced by the electron and positron distribution functions.  The observationally inferred value of \neff{} is affected by the electron/positron number densities and out-of-equilibrium entropy transfer between the neutrinos and the plasma.  We find that electron-positron annihilation freeze-out occurs at a sufficiently late time that there should be negligible effects on BBN yields and \neff.

Out-of-equilibrium effects are small ($\Delta \eta \sim 10^{-6}$) at $T \approx 24~{\rm keV}$, yet the BBN yields are set before this time, at higher plasma temperatures when $\Delta \eta$ is many orders of magnitude smaller.  (see, {\it  e.g.}, Ref.\ \cite{burst} for a discussion of BBN with out-of-equilibrium Boltzmann neutrino transport.)  As a result, we expect the BBN yields calculated with the standard assumption of thermal and chemical equilibrium in the electron/positron distributions to be insensitive to the loss of local chemical equilibrium observed in this calculation.

The value of $\neff = 3$ is predicated on a number of simplifying assumptions \cite{gf17} that neglect a number of effects.  Out-of-equilibrium scattering between neutrinos and electron/positrons distorts neutrino spectra, transfers entropy from the plasma to the neutrinos, and increases the total entropy of the universe \cite{burst}.  Non-zero electron/positron mass and finite-temperature QED affect the equation of state for electrons, positrons, and photons in the plasma.  Together, these effects increase \neff{} by $\Delta \neff \equiv \neff - 3 \approx 0.046$ \cite{burst,man05,dp16,byr15}.  The initial non-zero electron/positron degeneracy parameters are responsible for a negligible increase, $\Delta \neff \approx 10^{-19}$, due to a higher initial plasma entropy.  On the other hand, the non-zero final number density of electrons (and negligible number density of positrons) results in a negligible decrease, $\Delta \neff \approx - 10^{-7}$, because some of the entropy remains in the electron seas at recombination.

\section{Discussion and Conclusions} \label{sec:conclusion}

In this work, we found that the electrons and positrons remain in local chemical equilibrium throughout the BBN epoch and that this chemical equilibrium persists until the plasma temperature is approximately $16~{\rm keV}$, when the electron annihilation rate drops below the Hubble rate.  This annihilation freeze-out occurs after the BBN yields have been set and when the number densities of electrons and positrons are sufficiently low such that the value of \neff{} is unaffected.  To reach this conclusion, we self-consistently calculated the collision integral for electron-positron annihilation using the Klein-Nishina amplitude and the appropriate Fermi-blocking and Bose-enhancing statistical factors to take into account out-of-equilibrium effects.

The currently accepted value of \neff{} is predicated on the complete annihilation of the electrons and positrons from thermal number densities that are nearly equal to the photon number density to essentially zero \cite{gf17}.  The slight changes in the degeneracy parameters of electrons and positrons when the chemical equilibrium approximation is abandoned does not change this conclusion -- so as long as the standard cosmological model is assumed, annihilation freeze-out should have a negligible effect on the value of \neff.

BBN calculations assume chemical equilibrium between the electrons and positrons with the photons to model the creation/annihilation process ($e^-e^+ \rightleftharpoons \gamma \gamma$) \cite{kawano}.  We found that this approximation holds well throughout the BBN epoch, so BBN calculations in the standard cosmological model are not affected by the specifics of the annihilation freeze-out process.

BSM physics or non-standard cosmological models may push the annihilation freeze-out earlier (or nuclear freeze-out earlier) so that they may influence these cosmological observables.  Two possibilities that would require BSM physics to alter the standard cosmological model are a higher baryon-to-photon ratio during the BBN epoch and a higher effective electron mass.

While the baryon-to-photon ratio is directly measured from the CMB, this value is indicative of the baryon-to-photon ratio in the epochs immediately preceding photon decoupling, not necessarily at earlier epochs \cite{gfkp15}.  Although the deuterium and helium BBN yields and observations appear to form a concordance between the baryon-to-photon ratio at recombination and the BBN epochs, the lithium problem may suggest a more complicated evolution of this quantity.  A higher photon-to-baryon ratio during the BBN epoch would increase the annihilation freeze-out temperature, albeit by a slight amount unless it is many orders of magnitude larger.  
This would drastically alter the BBN yields, but also require a late generation of entropy to reduce the baryon-to-photon ratio to the CMB measured value \cite{fkk}.  Such an event would be a signal of BSM physics, and such a significant entropy generation would be strongly constrained by \neff{} measurements \cite{fkk} and possibly cascade nucleosynthesis that would determine the BBN yields \cite{cascade}.

Non-standard interactions between electrons and positrons and the plasma could increase the effective mass of the electrons and positrons \cite{gf17}.  A higher effective in-medium electron mass would cause the precipitous decline in electron and positron numbers seen in Figure \ref{fig:number_densities} to occur at a higher temperature, in turn, resulting in the strong decrease of $\mathcal{R}_{e^\pm}$ seen in Figure \ref{fig:annihilation_rate} occuring at a higher temperature as well.  The result is a higher annihilation freeze-out temperature.  The increase in the effective electron/positron mass due to finite temperature QED corrections is too small to significantly change the annihilation freeze-out temperature, so BSM interactions, and likely particles, would be required to produce such an effect, which brings up the same interesting questions about the effects on BBN yields and \neff.

In conclusion, we have computed the collision integral for electron-positron annihilation throughout the weak decoupling and BBN epochs of the early universe.  We find that electrons and positrons fall out of local chemical equilibrium with the photons at a sufficiently late time such that there will be no signature of this freeze-out on \neff{} or BBN yields.  However, one should be cautioned that non-standard cosmologies may result in non-trivial consequences of this annihilation freeze-out.

\acknowledgements
We'd like to thank G.\ Fuller for useful discussions.  CK and LT acknowledge support from NSF Grant PHY-1812383.  EG acknowledges support from NSF grant PHY-1630782 and the Heising-Simons Foundation, grant 2017-228.

\bibliographystyle{apsrev4-1}
\bibliography{refs}

\begin{thebibliography}{24}%
\makeatletter
\providecommand \@ifxundefined [1]{%
 \@ifx{#1\undefined}
}%
\providecommand \@ifnum [1]{%
 \ifnum #1\expandafter \@firstoftwo
 \else \expandafter \@secondoftwo
 \fi
}%
\providecommand \@ifx [1]{%
 \ifx #1\expandafter \@firstoftwo
 \else \expandafter \@secondoftwo
 \fi
}%
\providecommand \natexlab [1]{#1}%
\providecommand \enquote  [1]{``#1''}%
\providecommand \bibnamefont  [1]{#1}%
\providecommand \bibfnamefont [1]{#1}%
\providecommand \citenamefont [1]{#1}%
\providecommand \href@noop [0]{\@secondoftwo}%
\providecommand \href [0]{\begingroup \@sanitize@url \@href}%
\providecommand \@href[1]{\@@startlink{#1}\@@href}%
\providecommand \@@href[1]{\endgroup#1\@@endlink}%
\providecommand \@sanitize@url [0]{\catcode `\\12\catcode `\$12\catcode
  `\&12\catcode `\#12\catcode `\^12\catcode `\_12\catcode `\%12\relax}%
\providecommand \@@startlink[1]{}%
\providecommand \@@endlink[0]{}%
\providecommand \url  [0]{\begingroup\@sanitize@url \@url }%
\providecommand \@url [1]{\endgroup\@href {#1}{\urlprefix }}%
\providecommand \urlprefix  [0]{URL }%
\providecommand \Eprint [0]{\href }%
\providecommand \doibase [0]{http://dx.doi.org/}%
\providecommand \selectlanguage [0]{\@gobble}%
\providecommand \bibinfo  [0]{\@secondoftwo}%
\providecommand \bibfield  [0]{\@secondoftwo}%
\providecommand \translation [1]{[#1]}%
\providecommand \BibitemOpen [0]{}%
\providecommand \bibitemStop [0]{}%
\providecommand \bibitemNoStop [0]{.\EOS\space}%
\providecommand \EOS [0]{\spacefactor3000\relax}%
\providecommand \BibitemShut  [1]{\csname bibitem#1\endcsname}%
\let\auto@bib@innerbib\@empty
\bibitem [{\citenamefont {Grohs}\ \emph {et~al.}(2016)\citenamefont {Grohs},
  \citenamefont {Fuller}, \citenamefont {Kishimoto}, \citenamefont {Paris},\
  and\ \citenamefont {Vlasenko}}]{burst}%
  \BibitemOpen
  \bibfield  {author} {\bibinfo {author} {\bibfnamefont {E.}~\bibnamefont
  {Grohs}}, \bibinfo {author} {\bibfnamefont {G.~M.}\ \bibnamefont {Fuller}},
  \bibinfo {author} {\bibfnamefont {C.~T.}\ \bibnamefont {Kishimoto}}, \bibinfo
  {author} {\bibfnamefont {M.~W.}\ \bibnamefont {Paris}}, \ and\ \bibinfo
  {author} {\bibfnamefont {A.}~\bibnamefont {Vlasenko}},\ }\href@noop {}
  {\bibfield  {journal} {\bibinfo  {journal} {Phys.\ Rev.\ D}\ }\textbf
  {\bibinfo {volume} {93}},\ \bibinfo {pages} {083522} (\bibinfo {year}
  {2016})},\ \Eprint {http://arxiv.org/abs/arXiv:1512.02205} {arXiv:1512.02205}
  \BibitemShut {NoStop}%
\bibitem [{\citenamefont {de~Salas}\ and\ \citenamefont {Pastor}(2016)}]{dp16}%
  \BibitemOpen
  \bibfield  {author} {\bibinfo {author} {\bibfnamefont {P.~F.}\ \bibnamefont
  {de~Salas}}\ and\ \bibinfo {author} {\bibfnamefont {S.}~\bibnamefont
  {Pastor}},\ }\href@noop {} {\bibfield  {journal} {\bibinfo  {journal} {J.\
  Cosmol.\ Astropart.\ Phys.}\ }\textbf {\bibinfo {volume} {07}},\ \bibinfo
  {pages} {051} (\bibinfo {year} {2016})},\ \Eprint
  {http://arxiv.org/abs/arXiv:1606.06986} {arXiv:1606.06986} \BibitemShut
  {NoStop}%
\bibitem [{\citenamefont {Mangano}\ \emph {et~al.}(2005)\citenamefont
  {Mangano}, \citenamefont {Miele}, \citenamefont {Pastor}, \citenamefont
  {Pinto},\ and\ \citenamefont {Ofelia~Pisanti}}]{man05}%
  \BibitemOpen
  \bibfield  {author} {\bibinfo {author} {\bibfnamefont {G.}~\bibnamefont
  {Mangano}}, \bibinfo {author} {\bibfnamefont {G.}~\bibnamefont {Miele}},
  \bibinfo {author} {\bibfnamefont {S.}~\bibnamefont {Pastor}}, \bibinfo
  {author} {\bibfnamefont {T.}~\bibnamefont {Pinto}}, \ and\ \bibinfo {author}
  {\bibfnamefont {P.~D.~S.}\ \bibnamefont {Ofelia~Pisanti}},\ }\href@noop {}
  {\bibfield  {journal} {\bibinfo  {journal} {Nucl.\ Phys.\ B}\ }\textbf
  {\bibinfo {volume} {729}},\ \bibinfo {pages} {221} (\bibinfo {year}
  {2005})},\ \Eprint {http://arxiv.org/abs/hep-ph/0506164} {hep-ph/0506164}
  \BibitemShut {NoStop}%
\bibitem [{\citenamefont {Dolgov}\ \emph {et~al.}(2002)\citenamefont {Dolgov},
  \citenamefont {Hansen}, \citenamefont {Pastor}, \citenamefont {Petcov},
  \citenamefont {Raffelt},\ and\ \citenamefont {Semikoz}}]{dhpprs02}%
  \BibitemOpen
  \bibfield  {author} {\bibinfo {author} {\bibfnamefont {A.~D.}\ \bibnamefont
  {Dolgov}}, \bibinfo {author} {\bibfnamefont {S.~H.}\ \bibnamefont {Hansen}},
  \bibinfo {author} {\bibfnamefont {S.}~\bibnamefont {Pastor}}, \bibinfo
  {author} {\bibfnamefont {S.~T.}\ \bibnamefont {Petcov}}, \bibinfo {author}
  {\bibfnamefont {G.~G.}\ \bibnamefont {Raffelt}}, \ and\ \bibinfo {author}
  {\bibfnamefont {D.~V.}\ \bibnamefont {Semikoz}},\ }\href@noop {} {\bibfield
  {journal} {\bibinfo  {journal} {Nucl.\ Phys.\ B}\ }\textbf {\bibinfo {volume}
  {632}},\ \bibinfo {pages} {363} (\bibinfo {year} {2002})},\ \Eprint
  {http://arxiv.org/abs/hep-ph/0201287} {hep-ph/0201287} \BibitemShut {NoStop}%
\bibitem [{\citenamefont {Abazajian}\ \emph {et~al.}(2016)\citenamefont
  {Abazajian} \emph {et~al.}}]{CMBS4}%
  \BibitemOpen
  \bibfield  {author} {\bibinfo {author} {\bibfnamefont {K.~N.}\ \bibnamefont
  {Abazajian}} \emph {et~al.} (\bibinfo {collaboration} {CMB-S4}),\ }\href@noop
  {} {\  (\bibinfo {year} {2016})},\ \Eprint
  {http://arxiv.org/abs/arXiv:1610.02743} {arXiv:1610.02743} \BibitemShut
  {NoStop}%
\bibitem [{\citenamefont {Maiolino}\ \emph {et~al.}(2013)\citenamefont
  {Maiolino} \emph {et~al.}}]{EELT-HIRES}%
  \BibitemOpen
  \bibfield  {author} {\bibinfo {author} {\bibfnamefont {R.}~\bibnamefont
  {Maiolino}} \emph {et~al.},\ }\href@noop {} {\  (\bibinfo {year} {2013})},\
  \Eprint {http://arxiv.org/abs/arXiv:1310.3163} {arXiv:1310.3163} \BibitemShut
  {NoStop}%
\bibitem [{\citenamefont {Grohs}\ \emph {et~al.}(2015)\citenamefont {Grohs},
  \citenamefont {Fuller}, \citenamefont {Kishimoto},\ and\ \citenamefont
  {Paris}}]{gfkp15}%
  \BibitemOpen
  \bibfield  {author} {\bibinfo {author} {\bibfnamefont {E.}~\bibnamefont
  {Grohs}}, \bibinfo {author} {\bibfnamefont {G.~M.}\ \bibnamefont {Fuller}},
  \bibinfo {author} {\bibfnamefont {C.~T.}\ \bibnamefont {Kishimoto}}, \ and\
  \bibinfo {author} {\bibfnamefont {M.~W.}\ \bibnamefont {Paris}},\ }\href@noop
  {} {\bibfield  {journal} {\bibinfo  {journal} {J.\ Cosmol.\ Astropart.\
  Phys.}\ }\textbf {\bibinfo {volume} {05}},\ \bibinfo {pages} {017} (\bibinfo
  {year} {2015})}\BibitemShut {NoStop}%
\bibitem [{\citenamefont {Chu}\ \emph {et~al.}(2018)\citenamefont {Chu},
  \citenamefont {Dasgupta}, \citenamefont {Dentler}, \citenamefont {Kopp},\
  and\ \citenamefont {Saviano}}]{cddks18}%
  \BibitemOpen
  \bibfield  {author} {\bibinfo {author} {\bibfnamefont {X.}~\bibnamefont
  {Chu}}, \bibinfo {author} {\bibfnamefont {B.}~\bibnamefont {Dasgupta}},
  \bibinfo {author} {\bibfnamefont {M.}~\bibnamefont {Dentler}}, \bibinfo
  {author} {\bibfnamefont {J.}~\bibnamefont {Kopp}}, \ and\ \bibinfo {author}
  {\bibfnamefont {N.}~\bibnamefont {Saviano}},\ }\href@noop {} {\bibfield
  {journal} {\bibinfo  {journal} {J.\ Cosmol.\ Astropart.\ Phys.}\ }\textbf
  {\bibinfo {volume} {2018}},\ \bibinfo {pages} {049} (\bibinfo {year}
  {2018})},\ \Eprint {http://arxiv.org/abs/arXiv:1806.10629} {arXiv:1806.10629}
  \BibitemShut {NoStop}%
\bibitem [{\citenamefont {Fradette}\ \emph {et~al.}(2014)\citenamefont
  {Fradette}, \citenamefont {Pospelov}, \citenamefont {Pradler},\ and\
  \citenamefont {Ritz}}]{fppr14}%
  \BibitemOpen
  \bibfield  {author} {\bibinfo {author} {\bibfnamefont {A.}~\bibnamefont
  {Fradette}}, \bibinfo {author} {\bibfnamefont {M.}~\bibnamefont {Pospelov}},
  \bibinfo {author} {\bibfnamefont {J.}~\bibnamefont {Pradler}}, \ and\
  \bibinfo {author} {\bibfnamefont {A.}~\bibnamefont {Ritz}},\ }\href@noop {}
  {\bibfield  {journal} {\bibinfo  {journal} {Phys.\ Rev.\ D}\ }\textbf
  {\bibinfo {volume} {90}},\ \bibinfo {pages} {035022} (\bibinfo {year}
  {2014})},\ \Eprint {http://arxiv.org/abs/arXiv:1407.0993} {arXiv:1407.0993}
  \BibitemShut {NoStop}%
\bibitem [{\citenamefont {Berger}\ \emph {et~al.}(2016)\citenamefont {Berger},
  \citenamefont {Jedamzik},\ and\ \citenamefont {Walker}}]{bjw16}%
  \BibitemOpen
  \bibfield  {author} {\bibinfo {author} {\bibfnamefont {J.}~\bibnamefont
  {Berger}}, \bibinfo {author} {\bibfnamefont {K.}~\bibnamefont {Jedamzik}}, \
  and\ \bibinfo {author} {\bibfnamefont {D.~G.~E.}\ \bibnamefont {Walker}},\
  }\href@noop {} {\bibfield  {journal} {\bibinfo  {journal} {J.\ Cosmol.\
  Astropart.\ Phys.}\ }\textbf {\bibinfo {volume} {2016}},\ \bibinfo {pages}
  {032} (\bibinfo {year} {2016})},\ \Eprint
  {http://arxiv.org/abs/arXiv:1605.07195} {arXiv:1605.07195} \BibitemShut
  {NoStop}%
\bibitem [{\citenamefont {McDermott}(2016)}]{mcd18}%
  \BibitemOpen
  \bibfield  {author} {\bibinfo {author} {\bibfnamefont {S.~D.}\ \bibnamefont
  {McDermott}},\ }\href@noop {} {\bibfield  {journal} {\bibinfo  {journal}
  {Phys.\ Rev.\ Lett.}\ }\textbf {\bibinfo {volume} {120}},\ \bibinfo {pages}
  {221806} (\bibinfo {year} {2016})},\ \Eprint
  {http://arxiv.org/abs/arXiv:1711.00857} {arXiv:1711.00857} \BibitemShut
  {NoStop}%
\bibitem [{\citenamefont {Jedamzik}\ and\ \citenamefont
  {Pospelov}(2009)}]{jp09}%
  \BibitemOpen
  \bibfield  {author} {\bibinfo {author} {\bibfnamefont {K.}~\bibnamefont
  {Jedamzik}}\ and\ \bibinfo {author} {\bibfnamefont {M.}~\bibnamefont
  {Pospelov}},\ }\href@noop {} {\bibfield  {journal} {\bibinfo  {journal} {New
  J.\ Phys.}\ }\textbf {\bibinfo {volume} {11}},\ \bibinfo {pages} {105028}
  (\bibinfo {year} {2009})},\ \Eprint {http://arxiv.org/abs/arXiv:0906.2087}
  {arXiv:0906.2087} \BibitemShut {NoStop}%
\bibitem [{\citenamefont {Kolb}\ and\ \citenamefont {Turner}(1990)}]{kt90}%
  \BibitemOpen
  \bibfield  {author} {\bibinfo {author} {\bibfnamefont {E.~W.}\ \bibnamefont
  {Kolb}}\ and\ \bibinfo {author} {\bibfnamefont {M.~S.}\ \bibnamefont
  {Turner}},\ }\href@noop {} {\emph {\bibinfo {title} {The Early Universe}}}\
  (\bibinfo  {publisher} {Addison-Wesley},\ \bibinfo {year} {1990})\BibitemShut
  {NoStop}%
\bibitem [{\citenamefont {{Birrell}}\ \emph {et~al.}(2015)\citenamefont
  {{Birrell}}, \citenamefont {{Yang}},\ and\ \citenamefont
  {{Rafelski}}}]{byr15}%
  \BibitemOpen
  \bibfield  {author} {\bibinfo {author} {\bibfnamefont {J.}~\bibnamefont
  {{Birrell}}}, \bibinfo {author} {\bibfnamefont {C.~T.}\ \bibnamefont
  {{Yang}}}, \ and\ \bibinfo {author} {\bibfnamefont {J.}~\bibnamefont
  {{Rafelski}}},\ }\href@noop {} {\bibfield  {journal} {\bibinfo  {journal}
  {Nucl.\ Phys.\ B}\ }\textbf {\bibinfo {volume} {890}},\ \bibinfo {pages}
  {481} (\bibinfo {year} {2015})},\ \Eprint
  {http://arxiv.org/abs/arXiv:1406.1759} {arXiv:1406.1759} \BibitemShut
  {NoStop}%
\bibitem [{\citenamefont {Dolgov}\ \emph {et~al.}(1997)\citenamefont {Dolgov},
  \citenamefont {Hansen},\ and\ \citenamefont {Semikoz}}]{dhs97}%
  \BibitemOpen
  \bibfield  {author} {\bibinfo {author} {\bibfnamefont {A.~D.}\ \bibnamefont
  {Dolgov}}, \bibinfo {author} {\bibfnamefont {S.~H.}\ \bibnamefont {Hansen}},
  \ and\ \bibinfo {author} {\bibfnamefont {D.~V.}\ \bibnamefont {Semikoz}},\
  }\href@noop {} {\bibfield  {journal} {\bibinfo  {journal} {Nucl.\ Phys.\ B}\
  }\textbf {\bibinfo {volume} {503}},\ \bibinfo {pages} {426} (\bibinfo {year}
  {1997})},\ \Eprint {http://arxiv.org/abs/hep-ph/9703315} {hep-ph/9703315}
  \BibitemShut {NoStop}%
\bibitem [{\citenamefont {Mangano}\ \emph {et~al.}(2002)\citenamefont
  {Mangano}, \citenamefont {Miele}, \citenamefont {Pastor},\ and\ \citenamefont
  {Peloso}}]{man02}%
  \BibitemOpen
  \bibfield  {author} {\bibinfo {author} {\bibfnamefont {G.}~\bibnamefont
  {Mangano}}, \bibinfo {author} {\bibfnamefont {G.}~\bibnamefont {Miele}},
  \bibinfo {author} {\bibfnamefont {S.}~\bibnamefont {Pastor}}, \ and\ \bibinfo
  {author} {\bibfnamefont {M.}~\bibnamefont {Peloso}},\ }\href@noop {}
  {\bibfield  {journal} {\bibinfo  {journal} {Phys.\ Lett.\ B}\ }\textbf
  {\bibinfo {volume} {534}},\ \bibinfo {pages} {8 } (\bibinfo {year}
  {2002})}\BibitemShut {NoStop}%
\bibitem [{\citenamefont {Grohs}\ \emph {et~al.}(2017)\citenamefont {Grohs},
  \citenamefont {Fuller}, \citenamefont {Kishimoto},\ and\ \citenamefont
  {Paris}}]{gfkp17}%
  \BibitemOpen
  \bibfield  {author} {\bibinfo {author} {\bibfnamefont {E.}~\bibnamefont
  {Grohs}}, \bibinfo {author} {\bibfnamefont {G.~M.}\ \bibnamefont {Fuller}},
  \bibinfo {author} {\bibfnamefont {C.~T.}\ \bibnamefont {Kishimoto}}, \ and\
  \bibinfo {author} {\bibfnamefont {M.~W.}\ \bibnamefont {Paris}},\ }\href@noop
  {} {\bibfield  {journal} {\bibinfo  {journal} {Phys.\ Rev.\ D}\ }\textbf
  {\bibinfo {volume} {95}},\ \bibinfo {pages} {063503} (\bibinfo {year}
  {2017})},\ \Eprint {http://arxiv.org/abs/arXiv:1612.01986} {arXiv:1612.01986}
  \BibitemShut {NoStop}%
\bibitem [{\citenamefont {Kawano}(1992)}]{kawano}%
  \BibitemOpen
  \bibfield  {author} {\bibinfo {author} {\bibfnamefont {L.}~\bibnamefont
  {Kawano}},\ }\href@noop {} {\enquote {\bibinfo {title} {Let's go: Early
  universe 2. primordial nucleosynthesis the computer way},}\ } (\bibinfo
  {year} {1992})\BibitemShut {NoStop}%
\bibitem [{\citenamefont {Grohs}\ and\ \citenamefont {Fuller}(2016)}]{gf16}%
  \BibitemOpen
  \bibfield  {author} {\bibinfo {author} {\bibfnamefont {E.}~\bibnamefont
  {Grohs}}\ and\ \bibinfo {author} {\bibfnamefont {G.~M.}\ \bibnamefont
  {Fuller}},\ }\href@noop {} {\bibfield  {journal} {\bibinfo  {journal} {Nucl.\
  Phys.\ B}\ }\textbf {\bibinfo {volume} {911}},\ \bibinfo {pages} {955}
  (\bibinfo {year} {2016})}\BibitemShut {NoStop}%
\bibitem [{\citenamefont {Ade}\ \emph {et~al.}(2016)\citenamefont {Ade} \emph
  {et~al.}}]{planck}%
  \BibitemOpen
  \bibfield  {author} {\bibinfo {author} {\bibfnamefont {P.~A.~R.}\
  \bibnamefont {Ade}} \emph {et~al.} (\bibinfo {collaboration} {Planck
  collaboration}),\ }\href@noop {} {\bibfield  {journal} {\bibinfo  {journal}
  {Astron.\ Astrophys.}\ }\textbf {\bibinfo {volume} {594}},\ \bibinfo {pages}
  {A13} (\bibinfo {year} {2016})},\ \Eprint
  {http://arxiv.org/abs/arXiv:1502.01589} {arXiv:1502.01589} \BibitemShut
  {NoStop}%
\bibitem [{\citenamefont {Grohs}\ and\ \citenamefont {Fuller}(2017)}]{gf17}%
  \BibitemOpen
  \bibfield  {author} {\bibinfo {author} {\bibfnamefont {E.}~\bibnamefont
  {Grohs}}\ and\ \bibinfo {author} {\bibfnamefont {G.~M.}\ \bibnamefont
  {Fuller}},\ }\href@noop {} {\bibfield  {journal} {\bibinfo  {journal} {Nucl.\
  Phys.\ B}\ }\textbf {\bibinfo {volume} {923}},\ \bibinfo {pages} {222 }
  (\bibinfo {year} {2017})}\BibitemShut {NoStop}%
\bibitem [{\citenamefont {Fuller}\ \emph {et~al.}(2011)\citenamefont {Fuller},
  \citenamefont {Kishimoto},\ and\ \citenamefont {Kusenko}}]{fkk}%
  \BibitemOpen
  \bibfield  {author} {\bibinfo {author} {\bibfnamefont {G.~M.}\ \bibnamefont
  {Fuller}}, \bibinfo {author} {\bibfnamefont {C.~T.}\ \bibnamefont
  {Kishimoto}}, \ and\ \bibinfo {author} {\bibfnamefont {A.}~\bibnamefont
  {Kusenko}},\ }\href@noop {} {\  (\bibinfo {year} {2011})},\ \Eprint
  {http://arxiv.org/abs/arXiv:1110.6479} {arXiv:1110.6479} \BibitemShut
  {NoStop}%
\bibitem [{\citenamefont {Jedamzik}(2006)}]{cascade}%
  \BibitemOpen
  \bibfield  {author} {\bibinfo {author} {\bibfnamefont {K.}~\bibnamefont
  {Jedamzik}},\ }\href@noop {} {\bibfield  {journal} {\bibinfo  {journal}
  {Phys.\ Rev.\ D}\ }\textbf {\bibinfo {volume} {74}},\ \bibinfo {pages}
  {103509} (\bibinfo {year} {2006})},\ \Eprint
  {http://arxiv.org/abs/astro-ph/0604251} {astro-ph/0604251} \BibitemShut
  {NoStop}%
\bibitem [{\citenamefont {Peskin}\ and\ \citenamefont
  {Schroeder}(1995)}]{ps95}%
  \BibitemOpen
  \bibfield  {author} {\bibinfo {author} {\bibfnamefont {M.~E.}\ \bibnamefont
  {Peskin}}\ and\ \bibinfo {author} {\bibfnamefont {D.~V.}\ \bibnamefont
  {Schroeder}},\ }\href@noop {} {\emph {\bibinfo {title} {An Introduction to
  Quantum Field Theory}}}\ (\bibinfo  {publisher} {Westview Press},\ \bibinfo
  {year} {1995})\BibitemShut {NoStop}%
\end{thebibliography}%

\appendix

\section{Equations of Motion} \label{sec:appa}

Throughout this work we have assumed that the electron and positron distributions are described by a Fermi-Dirac spectrum, parameterized by a temperature, $T$ and degeneracy parameter, $\eta = \mu / T$,
\begin{equation}
    f ( p; T, \eta ) = \frac{1}{e^{E_p / T - \eta}+1} ,
\end{equation}
with $E_p = \sqrt{p^2 + m_e^2}$, and $T$ is the common temperature of all particles in the plasma.  This should be an excellent assumption since electromagnetic scattering between the elements of the photon-electron-positron-baryon plasma occurs at a rate much greater than the Hubble rate throughout the calculation.    

To self-consistently determine the equations of motion for $T (t)$ and $\eta(t)$, we need to calculate the energy density, $\rho$, pressure, $P$, and number density, $n$, of electrons and positrons from the distribution functions:
\begin{align}
    \rho (T, \eta) & = \int_0^\infty  \frac{p^2 d p}{\pi^2} \, E_p  \,f(p; T, \eta)   \\
    P (T, \eta) & =  \int_0^\infty \frac{p^2 dp}{\pi^2} \,\frac{p^2}{3 E_p} \,f(p; T, \eta)  \\
    n (T, \eta) & =  \int_0^\infty \frac{p^2 dp}{\pi^2} \,f(p; T, \eta) 
\end{align}

These quantities can be expressed in terms of dimensionless functions by defining $\epsilon = p / T$ and $x = m_e / T$,
\begin{align}
    I_1 (x, \eta) & = \frac{\rho}{T^4} = \frac{1}{\pi^2} \int_0^\infty d \epsilon \, \epsilon^2 E_\epsilon \, f(\epsilon; x, \eta)  \\
    I_2 (x, \eta) & = \frac{P}{T^4} = \frac{1}{3 \pi^2} \int_0^\infty d \epsilon \, \frac{\epsilon^4}{E_\epsilon} \, f(\epsilon; x, \eta)  \\
    N(x, \eta) & = \frac{n}{T^3} = \frac{1}{\pi^2} \int_0^\infty d \epsilon \, \epsilon^2 \, f(\epsilon; x, \eta) ,
\end{align}
where $E_\epsilon = \sqrt{\epsilon^2 + x^2}$.  Further, define the partial derivatives of these functions
\begin{align}
    J_i (x, \eta) = - \frac{\partial I_i}{\partial x} \qquad K_i (x, \eta) = \frac{\partial I_i}{\partial \eta} \\
    L (x, \eta) = - \frac{\partial N}{\partial x} \qquad M(x, \eta) = \frac{\partial N}{\partial \eta}
\end{align}
with $i = 1, 2$.  Each of these functions are, by design, dimensionless and positive definite.

\begin{widetext}
\subsection{Chemical equilibrium}

In the chemical equilibrium solution, $\eta_{e^-} = \eta$ and $\eta_{e^+} = - \eta$.  The total entropy density is
\begin{equation}
    s = T^3 \left( I_1^- + I_2^- + I_1^+ + I_2^+ + \frac{4 \pi^2}{45} - \eta N^- + \eta N^+ \right) + s_\nu .
\end{equation}
In the preceding equation, the `$-$' superscript corresponds to using the electron degeneracy parameter as the appropriate argument for each function, {\it e.g.}, $I_1^- = I_1 (x, \eta)$, while the `$+$' superscript corresponds to using the positron degeneracy parameter, {\it e.g.}, $I_1^+ = I_1 (x, -\eta)$.  While the entropy in the neutrino seas change through weak interactions beyond weak decoupling, it is a small effect that does not impact our results.  We therefore assume that neutrinos are decoupled from the plasma, so that the total entropy in the neutrino seas in a comoving volume, $s_\nu a^3$, is separately conserved.

The conservation of entropy in a comoving volume produces the differential equation:
\begin{align}
    & \left[ \left( I_1^- + I_2^- + I_1^+ + I_2^+ - \eta(N^- - N^+) + \frac{4\pi^2}{45} \right) \frac{3}{T} + \left( J_1^- + J_2^- + J_1^+ + J_2^+ - \eta (L^- - L^+) \right) \frac{m_e}{T^2} \right] \frac{dT}{dt} \nonumber \\
    & \qquad + \left[ K_1^- + K_2^- - K_1^+ - K_2^+ - (N^- - N^+) - \eta (M^- + M^+) \right] \frac{d \eta}{dt} \nonumber \\
    & \qquad \qquad + \left[ I_1^- + I_2^- + I_1^+ + I_2^+ - \eta (N^- - N^+) + \frac{4 \pi^2}{45} \right] \frac{da}{dt} = 0  .
    \label{eq:entropy_chemeq}
\end{align}

The evolution of the quantity $(n_{e^-} - n_{e^+}) a^3$, Eq.\ (\ref{eq:dndt_weak2}), yields the differential equation:
\begin{equation}
    \left[ \left( N^- - N^+ \right) \frac{3}{T} + \left( L^- - L^+ \right) \frac{m_e}{T^2}  \right] \frac{d T}{d t} + \left( M^- + M^+ \right)  \frac{d \eta}{d t} + \left( N^- - N^+ \right) \frac{3}{a} \frac{d a}{d t} = \frac{n_b a^3}{T^3 a^3} \frac{d Y_e}{d t} .
\end{equation}
The quantity, $n_b a^3$, is the number of baryons in a comoving volume is constant and equal to its initial value.

The evolution of $Y_e$ due to weak interactions, Eqs.\ (\ref{eq:weak_interaction1}-\ref{eq:weak_interaction3}) is 
\begin{equation}
    \frac{d Y_e}{d t} = (1-Y_e) (\lambda_{\nu_e n} + \lambda_{e^+ n} + \lambda_n) - Y_e (\lambda_{e^- p}+\lambda_{\bar{\nu}_e p}+\lambda_{\bar{\nu}_e e^- p}) .
    \label{eq:dyedt}
\end{equation}
Each of the weak rates are integrals over the electron distributions \cite{gf16}, so they depend on $T$ and $\eta$.  

The evolution of the scale factor is given by the Friedmann equation,
\begin{equation}
    \frac{1}{a} \frac{d a}{d t} = \sqrt{\frac{8 \pi}{3 m_{pl}^2}} \left[ \left( I_1^- + I_1^+ + \frac{\pi^2}{15} \right) T^4 + \frac{7 \pi^2}{40} T_{\rm cm}^4 , \right]^{1/2}
    \label{eq:friedmann}
\end{equation}
where $T_{\rm cm} = T_i (a_i / a)$, and we have neglected the baryon energy density.

These four coupled differential equations, Eqs.\ (\ref{eq:entropy_chemeq}-\ref{eq:friedmann}) define the evolution of the four dependent variables, $T$, $\eta$, $Y_e$ and $a$ using the initial conditions discussed in Section \ref{sec:chemeq}.

It should be noted, however, that $Y_e$ can be calculated directly (without a differential equation) from
\begin{equation}
    Y_e = \frac{T^3 a^3 (N^- - N^+)}{n_b a^3} .
\end{equation}
We chose to treat $Y_e$ as a dynamical variable with its own differential equation, but confirm that the above relationship between $Y_e$, $T$, $a$, and $\eta$ hold.

\subsection{Out of chemical equilibrium}

When we no longer assume chemical equilibrium, we are left with two independent degeneracy parameters, $\eta_{e^-} = \eta^-$ and $\eta_{e^+} = \eta^+$.  The total entropy density is 

\begin{equation}
    s = T^3 \left( I_1^- + I_2^- + I_1^+ + I_2^+ + \frac{4 \pi^2}{45} - \eta^- N^- - \eta^+ N^+ \right) + s_\nu .
\end{equation}
We use the same notation as previously used in this Appendix, where `$-$' corresponds to electrons, {\it e.g.}, $I_1^- = I_1 (x, \eta^-)$, and `$+$' corresponds to positrons, {\it e.g.}, $I_1^+ = I_1 (x, \eta^+)$.  The conservation of entropy in a comoving volume yields
\begin{align}
    & \left[ \left( I_1^- + I_2^- + I_1^+ + I_2^+ - \eta^- N^- - \eta^+ N^+ + \frac{4\pi^2}{45} \right) \frac{3}{T} + \left( J_1^- + J_2^- + J_1^+ + J_2^+ - \eta^- L^- - \eta^+ L^+ \right) \frac{m_e}{T^2} \right] \frac{dT}{dt} \nonumber \\
    & \qquad + \left( K_1^- + K_2^- - N^- - \eta^- M^-\right) \frac{d \eta^-}{dt} + \left( K_1^+ + K_2^+ - N^+ - \eta^+ M^+ \right) \frac{d \eta^+}{dt} \nonumber \\
    & \qquad \qquad + \left( I_1^- + I_2^- + I_1^+ + I_2^+ - \eta^- N^- - \eta^+ N^+ + \frac{4 \pi^2}{45} \right) \frac{da}{dt} = 0  .
    \label{eq:entropy_nochemeq}
\end{align}

The most easily distinguished difference between this equation and the corresponding one in the equilibrium scenario, Eq.\ (\ref{eq:entropy_chemeq}), are the separate degeneracy parameters, $\eta^-$ and $\eta^+$, and their derivatives.  

Likewise, in this scenario, the evolution of $(n_{e^-} - n_{e^+}) a^3$ yields the differential equation:
\begin{equation}
    \left[ \left( N^- - N^+ \right) \frac{3}{T} + \left( L^- - L^+ \right) \frac{m_e}{T^2}  \right] \frac{d T}{d t} + M^-\frac{d \eta^-}{d t} - M^+\frac{d \eta^+}{d t}   + \left( N^- - N^+ \right) \frac{3}{a} \frac{d a}{d t} = \frac{n_b a^3}{T^3 a^3} \frac{d Y_e}{d t} .
\end{equation}

The evolution of $Y_e$ and $a$, Eqs.\ (\ref{eq:dyedt},\ref{eq:friedmann}), remain unchanged.  However, it should be noted that the weak rates depend on both $\eta^-$ and $\eta^+$.

Now that there are two independent degeneracy parameters, we require an extra independent differential equation.  We introduced the evolution of $(n_{e^+} + n_{e^-}) a^3$, Eq.\ (\ref{eq:dndt_sum}), which results in the differential equation:
\begin{align}
    & \left[ \left( N^- + N^+ \right) \frac{3}{T} + \left( L^- + L^+ \right) \frac{m_e}{T^2} \right] \frac{d T}{d t} + M^- \frac{d \eta^-}{d t} + M^+ \frac{d \eta^+}{d t} + \left( N^- - N^+ \right) \frac{3}{a} \frac{d a}{d t} \nonumber \\
    & \qquad = \frac{n_b a^3}{T^3 a^3} \left[ (1 - Y_e) ( \lambda_{\nu_e n} - \lambda_{e^+ n} + \lambda_n ) + Y_e (- \lambda_{e^- p} + \lambda_{\bar\nu_e p} - \lambda_{\bar\nu_e e^- p} ) + 2  \frac{\mathcal{N} a^3}{n_b a^3} \right] .
    \label{eq:newchemeq_eqn}
\end{align}
As before, the $n_b a^3$ is left explicitly uncancelled in the expressions above because it is constant.  The net electron-positron annihilation rate, $\mathcal{N}$, is defined in Eq.\ (\ref{eq:dndt_netrate}).

The five coupled differential equations, Eqs.\ (\ref{eq:dyedt},\ref{eq:friedmann},\ref{eq:entropy_nochemeq}-\ref{eq:newchemeq_eqn}) define the evolution of five dependent variables, $T$, $\eta^-$, $\eta^+$, $Y_e$ and $a$, whose initial conditions and solutions are discussed in Section \ref{sec:nochemeq}.

\section{Simplification of the Collision Integral} \label{sec:appb}
This appendix details the reduction of the collision integral for electron-positron annihilation from a nine-dimensional integral to three-dimensional in a manner similar to the appendix in Ref.\ \cite{burst}.  For the annihilation process, $e^-  + e^+  \rightleftharpoons \gamma  + \gamma$, the Klein-Nishina amplitude is \cite{ps95}
\begin{equation}
    \langle \vert \mathcal{M} \vert^2 \rangle = 2 e^4 \left[ \frac{P \cdot K_2}{P \cdot K_1} + \frac{P \cdot K_1}{P \cdot K_2} + 2 m_e^2 \left( \frac{1}{P \cdot K_1} + \frac{1}{P \cdot K_2} \right) - m_e^4 \left( \frac{1}{P \cdot K_1} + \frac{1}{P \cdot K_2} \right)^2 \right] \equiv M  ,
    \label{eq:amplitude}
\end{equation}
where $e$ is the elementary charge, $m_e$ is the electron mass, $P$ and $Q$ are the four-momenta of the electron and positron, respectively, and $K_1$ and $K_2$ are the photon four-momenta. We define the four-momenta of the electron as $P = (E, \v{p})$ with particle energy $E$ and three-momentum \v{p} such that $P \cdot P = E^2 - p^2 = m_e^2$, with $p = \vert \v{p} \vert$, and the same convention for positrons.  Likewise, for the photon, $K = (k, \v{k})$ with photon three-momentum \v{k} and energy $k = \vert \v{k} \vert$.

The collision integral is
\begin{equation}
    C = \frac{1}{2 E_p} \int \frac{d^3 q}{(2 \pi)^3 2 E_q} \frac{d^3 k_1}{(2 \pi)^3 2 k_1} \frac{d^3 k_2}{(2 \pi)^3 2 k_2} (2 \pi)^4 \delta^{(4)} (P + Q - K_1 - K_2) M ( P \cdot K_1, P \cdot K_2 ) F(E_p, E_q, k_1, k_2) ,
\end{equation}
and 
\begin{align}
    F(E_p, E_q, k_1, k_2) & \equiv F_c - F_a \nonumber \\ 
    & = f_\gamma(k_1)f_\gamma(k_2)[1-f_{e^-}(E_p)][1-f_{e^+}(E_q)] - f_{e^-}(E_p)f_{e^+}(E_q)[1+f_\gamma(k_1)][1+f_\gamma(k_2)] 
\end{align}
is the statistical factor for the creation ($F_c$) and annihilation ($F_a$) of electrons and positrons with the appropriate Fermi blocking and Bose enhancement terms. For the case $F=F_a$, we define $C_a$ as the collision integral for solely the annihilation pathway ($e^-e^+\rightarrow\gamma\gamma$) and $F = F_c$ defines $C_c$ for the creation pathway ($\gamma \gamma \rightarrow e^- e^+$).

Integrating over $\v{k}_2$ reduces the delta function to a one-dimensional energy-conserving delta function:
\begin{equation}
    C = \frac{1}{(2 \pi)^5 16 E_p} \left. \int \frac{d^3 q}{E_q} \int \frac{d^3 k_1}{k_1 k_2} \delta (E_p + E_q - k_1 - k_2) M(P \cdot K_1, P \cdot K_2) F(E_p, E_q, k_1, k_2) \right\vert_{k_2 = \vert \v{p} + \v{q} - \v{k}_1 \vert} ,
\end{equation}
where $K_2 = (k_2, \v{p} + \v{q} - \v{k}_1)$, and $k_2$ is no longer an independent variable of integration, but is related to the other integration variables through momentum conservation, and
\begin{equation}
    k_2^2 = \vert \v{p} + \v{q} \vert^2 + k_1^2 - 2 \vert \v{p} + \v{q} \vert k_1 \cos \theta_1,
\end{equation}
defining the new integration angle $\theta_1$ angle between the vectors $\v{k}_1$ and $\v{p}+\v{q}$.

We need to evaluate the four-dot products
\begin{align}
    P \cdot K_1 & = E_p k_1 - \v{p} \cdot \v{k}_1 \\
    P \cdot K_2 & = E_p k_2 - \v{p} \cdot (\v{p} + \v{q} - \v{k}_1) = E_p k_2 - p^2 - \v{p} \cdot \v{q} + \v{p} \cdot \v{k}_1 ,
\end{align}
which requires us to parameterize the dot products $\v{p} \cdot \v{q}$ and $\v{p} \cdot \v{k}_1$ in terms of the variables of integration.  First, define $\theta_q$ as the angle between \v{p} and \v{q} so that
\begin{align}
    \v{p} \cdot \v{q} & = p q \cos \theta_q  \\
    \vpq & = \left( p^2 + q^2 + 2 p q \cos \theta_q \right)^{1/2}.
\end{align}
The vector $\v{p}+\v{q}$ is in the plane generally defined by \v{p} and \v{q}, but $\v{k}_1$ is not necessarily in this plane.  This means $\v{k}_1$ is defined by both $\theta_1$, the angle between it and $\v{p}+\v{q}$, and an azimuthal angle $\phi_1$ measured out of this plane.  To determine $\v{p} \cdot \v{k}_1$, we need to introduce another angle, $\psi$, between \v{p} and $\v{p}+\v{q}$.  It follows that
\begin{align}
    \v{p} \cdot \v{k}_1 = p k_1 \left( \cos \psi \cos \theta_1 - \sin \psi \sin \theta_1 \cos \phi_1 \right) , \\
    {\rm with}~ \cos \psi = \frac{\v{p} \cdot (\v{p}+\v{q})}{p \vert \v{p}+\v{q} \vert} = \frac{p + q \cos \theta_q}{\sqrt{p^2 + q^2 + 2 p q \cos \theta_q}} .
    \label{eq:psi}
\end{align}

This parameterizes the amplitude in terms of almost all the integration variables:
\begin{equation}
    M(P \cdot K_1, P \cdot K_2) = M(p, q, \theta_q, k_1, \theta_1, \phi_1) .
\end{equation}
When considering $d^3 q = q^2 \, dq \, d \cos \theta_q \, d \phi_q$ and $d^3 k = k^2 \, dk \, d \cos \theta_1 \, d \phi_1$, the integrand is independent of $\phi_q$, so this integral can be performed which results in a factor of $2 \pi$, so
\begin{equation}
    C = \frac{1}{(2 \pi)^4 16 E_p} \int \frac{q^2 \, dq \, d \cos \theta_q}{E_q} \int \frac{k_1^2 \, d k_1 \, d \cos \theta_1 \, d \phi_1}{k_1 k_2} \delta(E_p + E_q - k_1 - k_2) M(p,q,\theta_q,k_1,\theta_1,\phi_1) F(E_p, E_q, k_1, k_2) .
\end{equation}

To complete the integral over $\theta_1$, we introduce the $u$ substitution,
\begin{equation}
    u^2 = k_2^2 \qquad \Rightarrow \qquad 2 k_2 \, du = - 2 \vert \v{p} + \v{q} \vert \, k_1 d \cos \theta_1 .
\end{equation}
This allows us to rewrite the colision integral as
\begin{equation}
    C = \frac{1}{(2 \pi)^4 16 E_p}  \int \frac{q^2 \, d q \, d \cos_q}{E_q \vert \v{p} + \v{q} \vert} \int d k_1 \, d \phi_1 \int_a^b d u \, \delta( E_p + E_q - k_1 - u) M(p, q, \theta_q, k_1, \theta_1, \phi_1) F(E_p, E_q, k_1, u) ,
\end{equation}
with 
\begin{align}
    a & = \left[ \vert \v{p} + \v{q} \vert^2 + k_1^2 - 2 \vert \v{p} + \v{q} \vert k_1 \right]^{1/2} = \left\vert \vert \v{p} + \v{q} \vert - k_1 \right\vert , \\
    b & = \left[ \vert \v{p} + \v{q} \vert^2 + k_1^2 + 2 \vert \v{p} + \v{q} \vert k_1 \right]^{1/2} =  \vert \v{p} + \v{q} \vert + k_1 .
\end{align}

The delta function is non-zero on the interval $[a, b]$ only if
\begin{equation}
    a < E_p + E_q - k_1 < b ,
\end{equation}
which will constrain the range of the other integrals once the delta function is resolved.  To determine these ranges we need to consider two cases:

{\it Case 1:} $k_1 < \vpq$.  The inequality has two parts, first:
\begin{align}
    & \vpq - k_1 < E_p + E_q - k_1 \nonumber \\
    & \quad \Rightarrow \quad \vpq < E_p + E_q .
\end{align}
This is always true because $\vpq < p + q$ (triangle inequality), and momenta are always less than their corresponding energy.  Further, the second inequality:
\begin{align}
    & E_p + E_q - k_1 < \vpq + k_1 \nonumber \\
    & \quad \Rightarrow \quad k_1 > \frac{1}{2} \left( E_p + E_q - \vpq \right) \equiv k_{\rm min}.
\end{align}
The right-hand side is always positive, as stated previously.  This reduces the range of the $k_1$ integral such that its integrand is non-zero.

{\it Case 2:} $k_1 > \vpq$.  This case differs from Case 1 only in the first inequality,
\begin{align}
    & k_1 - \vpq < E_p + E_q - k_1 \nonumber \\
    & \quad \Rightarrow \quad k_1 < \frac{1}{2} \left( E_p + E_q + \vpq \right) \equiv k_{\rm max} ,
\end{align}
and again, the second inequality yields $k_1 > k_{\rm min}$.  

Putting the cases together, we find that the delta function is non-zero when $k_{\rm min} < k_1 < k_{\rm max}$, so upon evaluating the $du$ integral the delta function fixes $\cos(\theta_1)$ in terms of other variables:
\begin{align}
    \cos \theta_1 & = \frac{\vpq^2 + k_1^2 - k_2^2}{2 k_1 \vpq} \nonumber \\
     & = \frac{\vpq^2 - (E_p + E_q)^2 + 2 k_1 (E_p + E_q)}{2 k_1 \vpq} .
     \label{eq:theta1}
\end{align}
The collision integral then becomes
\begin{equation}
    C = \frac{1}{(2 \pi)^4 16 E_p}  \int \frac{q^2 \, d q \, d \cos_q}{E_q \vert \v{p} + \v{q} \vert} \int_{k_{\rm min}}^{k_{\rm max}} d k_1\left[\int_0^{2\pi} d \phi_1 M(p, q, \theta_q, k_1, \theta_1, \phi_1)\right]F(E_p, E_q, k_1, k_2),
\end{equation}
where, once again, $k_2 = E_p + E_q - k_1$. Note that only $M$ depends on $\phi_1$, so we can perform the integral in the square brackets analytically. To simplify our notation, we define the function 
\begin{equation}
    \tilde{M} (p, q, \theta_q, k_1) = \frac{1}{(2 \pi) (2 e^4)} \int_0^{2 \pi} d \phi_1 \, M(p, q, \theta_q, k_1, \theta_1, \phi_1) .
\end{equation}

The amplitude, Eq.\ (\ref{eq:amplitude}), depends only on the four-dot products $P \cdot K_1$ and $P \cdot K_2$.  These dot products can be written in the form $A + B \cos \phi_1$.  The function $\tilde{M}$, above, can be analytically integrated as long as $A > B$.  These four-dot products can be shown to be positive definite, which proves that $A > B$, by considering the annihilation process in the center-of-momentum frame where the four-momenta can be written as $P = (E_p, \v{p})$, $Q = (E_p, - \v{p})$, $K_1 = (k, \v{k})$, and $K_2 = (k, - \v{k})$.
\begin{align}
    P \cdot K_1 & = k (E_p + p \cos \vartheta) > 0 \nonumber \\
    P \cdot K_2 & = k (E_p - p \cos \vartheta) > 0 ,
\end{align}
where $\vartheta$ is the angle between \v{p} and \v{k}.  This means that the individual terms in the amplitude are positive definite in any frame, and allows us to analytically perform the integral resulting in:
\begin{align}
    \tilde{M} (p, q, \theta_q, k_1) & = - 2 + (3 m_e^2 + E_p E_q - p q \cos \theta_q ) \left( \frac{1}{\sqrt{X^2 - Y^2}} + \frac{1}{\sqrt{Z^2 - Y^2}} \right) \nonumber \\
    & \qquad - m_e^4 \left[ \frac{X}{(X^2 - Y^2)^{3/2}} + \frac{Z}{(Z^2 - Y^2)^{3/2}} + \frac{2}{X+Z} \left( \frac{1}{\sqrt{X^2-Y^2}} + \frac{1}{\sqrt{Z^2-Y^2}} \right) \right]
\end{align}
\begin{align}
    X & = E_p k_1 - p k_1 \cos \psi  \cos \theta_1 \\
    Y & = p k_1 \sin \psi  \sin \theta_1 \\
    Z & = m_e^2 + E_p E_q - p q \cos \theta_q - X
\end{align}
$\psi$ and $\theta_1$ are defined in terms of $p$, $q$, $\theta_q$, and $k_1$ in Eqs.\ (\ref{eq:psi}) and (\ref{eq:theta1}).

The final form of the collision integral is therefore
\begin{equation}
    C = \frac{e^4}{(2 \pi)^3 8 E_p} \int_0^\infty \frac{q^2 \, dq}{E_q} \int_{-1}^1 \frac{d \cos \theta_q}{\vpq} \int_{k_{\rm min}}^{k_{\rm max}} d k_1\, \tilde{M} (p, q, \theta_q, k_1) F(E_p, E_q, k_1, E_p + E_q - k_1)  .
\end{equation}
This integral is calculated using Gauss-Legendre quadrature for the two inner integrals ($\cos \theta_q$, $k_1$) over finite ranges and using Gauss-Laguerre quadrature for the outer integral ($q$) over the infinite range.

\end{widetext}

\end{document}